\documentclass{ws-procs9x6-cpt16}
\def\etal {{\it et al.}}
\begin{document}

\title{General Electromagnetic Nonminimal Couplings\\
in the Dirac Equation}

\author{J.B.\ Araujo, Rodolfo Casana, and M.M.\ Ferreira, Jr.}

\address{Departamento de F\'{\i}sica, Universidade Federal do Maranh\~{a}o\\ 
Campus Universit\'{a}rio do Bacanga, S\~{a}o Lu\'{\i}s, Maranh\~{a}o
65080-805, Brazil}

\begin{abstract}
We examine a new class of CPT-even and dimension-five nonminimal
interactions between fermions and photons, deprived of higher-order
derivatives, yielding electric dipole moment and magnetic dipole
moment in the context of the Dirac equation. These couplings are
Lorentz-violating nonminimal structures, composed of a rank-2 tensor,
the electromagnetic tensor, and gamma matrices, being
addressed in its axial and non-axial hermitian versions. We use the
electron's anomalous magnetic and electric dipole moment measurements to reach upper bounds of $1$
part in $10^{11}$ and $10^{16}$ GeV$^{-1}$.
\end{abstract}

\bodymatter 

\phantom{}\vskip10pt\noindent
The Standard Model (SM) structure allows for C, P, and T violations
(and combinations) as long as CPT symmetry is preserved. Among the
most important tests for physics beyond the SM is the search for the
electric dipole moment (EDM) of particles in connection with the strong CP
problem.\cite{Axion,LeptonEDM} The EDM nonrelativistic interaction term
has the form $d(\boldsymbol{\sigma }\cdot \mathbf{E}),$ with $\mathbf{E}$, $\boldsymbol{\sigma }$ and $d$ as the electric field, the spin operator, and the
EDM modulus, respectively. This interaction violates both P and T symmetries, $P(%
\boldsymbol{\sigma }\cdot \mathbf{E})\rightarrow -(\boldsymbol{\sigma }\cdot
\mathbf{E})$, $T(\boldsymbol{\sigma }\cdot \mathbf{E})\rightarrow -(%
\boldsymbol{\sigma }\cdot \mathbf{E})$, but preserves C and CPT
symmetries. Lorentz-violating nonminimal couplings in the context of the
Dirac equation that engender EDM at tree level can be stringently
constrained by EDM data.\cite{Pospelov, FredeNM1,Jonas1}
Theories endowed with Lorentz violation are a possible route for physics beyond SM that
has been intensively studied in the latest years, mainly in the context of the
Standard-Model Extension (SME).\cite{SME,Sideral} Connections
between LV theories and EDM physics have been addressed in Ref.\ \refcite{Altarev}.

In this work, a new class of dimension-five nonminimal couplings is proposed
and examined with respect to the possible generation of an EDM 
and magnetic dipole moment (MDM),\cite{Jonas2} having the hermiticity as the main criterion to be observed.

Nonminimal couplings in the context of the Dirac equation can be stated by
modifying the usual covariant derivative.\cite{FredeNM1,Jonas1} For
the cases involving a rank-$2,$ dimensionless, LV tensor $T_{\mu \nu }$, a
first proposal is given by $D_{\mu }=\text{$\partial $}_{\mu }+ieA_{\mu
}+i\lambda _{1}T_{\mu \nu }F^{\nu \beta }\gamma ^{5}\gamma _{\beta }$. Its
hermitian version is $D_{\mu }=\text{$\partial $}_{\mu }+ieA_{\mu }-\frac{i}{%
2}\lambda _{1}\left( T_{\mu \nu }F^{\nu \beta }-T_{\beta \nu }F^{\nu }{}_{\mu}\right)
\gamma _{\beta }\gamma ^{5}$, which leads to the modified
Dirac equation
\begin{equation}
\left[ i\gamma ^{\mu }\partial _{\mu }-e\gamma ^{\mu }A_{\mu }-i\lambda
_{1}T_{\mu \nu }F_{\text{ \ }\beta }^{\nu }\sigma ^{\mu \beta }\gamma ^{5}-m%
\right] \Psi =0,  \label{DiracM1}
\end{equation}%
In momentum coordinates, the Dirac equation can be rewritten as $%
i\partial _{t}\Psi =\left[ \boldsymbol{\alpha }\cdot \boldsymbol{\pi }%
+eA_{0}+m\gamma ^{0}+H_{\text{LV}}\right] \Psi ,$ with $H_{\text{LV}%
}=i\lambda _{1}T_{\mu \nu }F^{\nu }{}_{\beta }\gamma ^{0}\sigma
^{\mu \beta }\gamma ^{5}$ being the LV piece of the hamiltonian. Using $%
F_{0j}=E^{j},F_{mn}=\epsilon _{mnp}B_{p}$, $\sigma ^{0j}=i\alpha ^{j}$ and $%
\sigma ^{ij}=\epsilon _{ijk}\Sigma ^{k}$, the nonrelativistic hamiltonian
for uniform fields is 
\begin{align}
H_{\text{NR}}& =H_{\text{Pauli}}+\frac{1}{2m}\left[ (\boldsymbol{\sigma }%
\cdot \boldsymbol{\pi )}H_{12}-H_{12}(\boldsymbol{\sigma }\cdot \boldsymbol{%
\pi })\right]   \notag \\
& -\lambda _{1}\left( T_{00}\boldsymbol{\sigma }\cdot \boldsymbol{E}+T_{0i}(%
\boldsymbol{\sigma }\times \boldsymbol{B})^{i}-T_{ij}E^{j}\sigma ^{i}\right)
, \label{HNR3}
\end{align}%
with $H_{\text{Pauli}}=\frac{1}{2m}\left[ \boldsymbol{\pi }{}^{2}-e%
\boldsymbol{\sigma }\cdot \boldsymbol{B}\right] +eA_{0},$ $H_{12}=i\lambda
_{1}(-\epsilon _{ijk}T_{i0}\sigma ^{k}E^{j}-T_{ii}\left( \boldsymbol{\sigma }%
\cdot \boldsymbol{B}\right) +T_{ij}\sigma ^{j}B^{i}).$ We notice that the
terms $\lambda _{1}T_{00}(\boldsymbol{\sigma }\cdot \boldsymbol{E)}$, $%
\lambda _{1}T_{ij}E^{j}\sigma ^{i}$ are able to generate an EDM 
for the electron at tree level, so that the LV coefficients $%
\lambda _{1}T_{00},$ $\lambda _{1}T_{ij}$ can be constrained by using 
EDM experimental data.\cite{Baron} The generation of MDM is unusual and
associated with the term, $\lambda _{1}T_{0i}(\sigma \times B)^{i}$, by an
unconventional interpretation: if we consider $\lambda _{1}T_{0i}(\sigma
\times B)^{i}=\lambda _{1}\boldsymbol{\tilde{\sigma}}\cdot \mathbf{B}%
=\lambda _{1}\tilde{\sigma}^{k}B^{k}$, where $\tilde{\sigma}^{k}=\epsilon
_{ijk}T_{0i}\sigma ^{j}$ is a kind of ``rotated" spin operator by the LV
background, we should be able to use the MDM experimental 
data\cite{Gabrielse} to impose bounds on the parameter $\lambda _{1}T_{0i}$ as well
(if a suitable experimental set up able to measure MDM in an orthogonal
direction to the applied field is found). Table I displays the obtained
bounds for this coupling, its non-axial version, and other relevant
cases.

\begin{table}
\tbl{List of the bounds (in GeV$^{-1}$ units) for the relevant couplings.}
{\begin{tabular}{@{}ccc@{}}\toprule
Coupling & EDM & MDM  \\\colrule
$i\lambda_{1}\bar{\Psi}T_{\mu\nu}F^{\nu}{}_{\beta}\sigma^{\mu\beta}\gamma^{5}\Psi$& $|\lambda_{1}T_{00}|\leq3.8\times10^{-16} $  & $|\lambda_{1}T_{0i}|\leq5.5\times10^{-11}$   \\
&$|\lambda_{1}T|\leq1.1\times10^{-15} $ & \\
$\lambda^{\prime }_{1}\bar{\Psi}T_{\mu\nu}F^{\nu}{}_{\beta}\sigma^{\mu\beta}\Psi$ & $|\lambda^{\prime }_{1}T_{i0}|\leq3.8\times10^{-16}$ &$|\lambda^{\prime }_{1}T|\leq3.5\times10^{-11}$  \\
$\lambda_{4}\bar{\Psi}(T_{\alpha\nu}F_{\mu\beta}-T_{\mu\beta}F_{\alpha\nu})\sigma^{\mu\beta}\sigma^{\alpha\nu}\Psi$& $|\lambda_{4}T_{0i}|\leq3.8\times10^{-16} $ & $|\lambda_{4}T_{ij}|\leq5.5\times10^{-11} $   \\\botrule
\end{tabular}} \label{aba:tbl1}
\end{table}

The non-axial version of this coupling is given by the following
extension to the covariant derivative: $D_{\mu }=\partial _{\mu }+ieA_{\mu }+%
\frac{\lambda _{1}^{\prime }}{2}\left( T_{\mu \nu }F^{\nu}{}_{\beta }-T_{\beta \nu }
F^{\nu }{}_{\mu }\right) \gamma ^{\beta }$, which
yields a hermitian contribution to the Dirac equation,
\begin{equation}
\left[ i\gamma ^{\mu }\partial _{\mu }-e\gamma ^{\mu }A_{\mu }+\lambda
_{1}T_{\mu \nu }F_{\text{ \ }\beta }^{\nu }\sigma ^{\mu \beta }-m\right]
\Psi =0,  \label{DiracM2}
\end{equation}%
implying the hamiltonian LV piece, $H_{\text{LV}}^{\prime }=-\lambda
_{1}^{\prime }T_{\mu \nu }F_{\text{ \ }\beta }^{\nu }\gamma ^{0}\sigma ^{\mu
\beta }$. The full nonrelativistic hamiltonian is
\begin{align}
H_{\text{NR}}& =H_{\text{Pauli}}+\frac{1}{2m}[\left( \boldsymbol{\sigma }%
\cdot \boldsymbol{\pi }\right) H_{12}^{\prime }-H_{12}^{\prime }\left(
\boldsymbol{\sigma }\cdot \boldsymbol{\pi }\right) ]  \notag \\
& +\lambda _{1}^{\prime }T_{i0}\left( \boldsymbol{\sigma }\times \boldsymbol{%
E}\right) ^{i}-\lambda _{1}^{\prime }T\left( \boldsymbol{\sigma }\cdot
\boldsymbol{B}\right) +\lambda _{1}^{\prime }T_{ij}\sigma ^{j}B^{i},
\label{HNR5}
\end{align}%
where $T=\text{Tr}(T_{ij})$ and $H_{12}^{\prime }=i\lambda _{1}^{\prime
}(T_{00}E^{i}\sigma ^{i}-T_{ij}E^{j}\sigma ^{i}+T_{0i}\epsilon _{ijk}\sigma
^{j}B^{k}).$ Clearly, the terms $T\left( \boldsymbol{\sigma }\cdot
\boldsymbol{B}\right) $\ and $T_{ij}\sigma ^{j}B^{i}$ are MDM contributions,
involving only the symmetric part of the tensor $T_{\mu \nu }$, whose
magnitude can be limited by the known experimental error 
in the MDM.\cite{Gabrielse} There is also an unconventional EDM contribution. Analogously
to the magnetic case, the term $\lambda _{1}^{\prime }T_{i0}\left(
\boldsymbol{\sigma }\times \boldsymbol{E}\right) ^{i}=\lambda _{1}^{\prime }%
\tilde{\sigma}^{k}E^{k}$ could be considered as EDM if we take $\tilde{\sigma%
}^{k}=\epsilon _{ijk}T_{0i}\sigma ^{j}$ as a kind of ``rotated" EDM.
Therefore, EDM and MDM experimental data can be used to constrain these
terms, as it is shown in Table I. An interesting discussion concerning the
overlapping of some bounds with others in the literature is found in 
Ref.\ \refcite{Jonas2}. For a discussion on the sidereal variations, to which the LV
parameters are subject, see Refs.\ \refcite{Jonas1,Sideral,Jonas2}.

The nonminimal couplings of Eqs.\ (\ref{DiracM1}) and (\ref{DiracM2}) enclose
two gamma matrices contracted with the background and the $F^{\alpha \beta }$
tensor. New possibilities of couplings arise when the tensors $T_{\mu \nu }$
and $F^{\alpha \beta }$ have no mutually contracted indices, leaving three
indices to be contracted with gamma matrices. These derivatives are
comprised in the general expression $D_{\mu }=\partial _{\mu }+ieA_{\mu
}+i\lambda _{3}T_{\{\alpha \nu }F_{\mu \beta \}}\gamma ^{\{\beta }\gamma
^{\alpha }\gamma ^{\nu \}},$ with the symbol $\{\}$ denoting possible
permutation of the indexes $\mu ,\nu ,\alpha ,\beta $. Among them there is
\begin{equation}
D_{\mu }=\partial _{\mu }+ieA_{\mu }+i\frac{\lambda _{4}}{8}\left( T_{\alpha
\nu }F_{\mu \beta }-T_{\mu \beta }F_{\alpha \nu }\right) \gamma ^{\beta
}\sigma ^{\alpha \nu }.
\end{equation}%
This coupling is distinct from the previous ones, leading to the Dirac
equation contribution: $[\lambda _{4}T^{\nu }{}_{\nu }F_{\mu \beta }\sigma
^{\mu \beta }+i\lambda _{4}\left( T_{\alpha \nu }F_{\mu \beta }-T_{\mu \beta
}F_{\alpha \nu }\right) \sigma ^{\mu \beta }\sigma ^{\alpha \nu }]/8.$ In
this case, the symmetric part of $T_{\mu \nu }$ yields the usual MDM
interaction, $\lambda _{4}T^{\nu }{}_{\nu }F_{\mu \beta }\bar{\Psi}\sigma
^{\mu \beta }\Psi $. Taking an antisymmetric $T_{\mu \nu }$, the modified
Dirac equation assumes the form $i\partial _{t}\Psi =\left[ \boldsymbol{%
\alpha }\cdot \boldsymbol{\pi }+eA_{0}+m\gamma ^{0}+H_{\text{LV4}}\right]
\Psi ,$ where the LV piece is $H_{\text{LV4}}=i\frac{\lambda _{4}}{8}\left[
T_{\alpha \nu }F_{\mu \beta }-T_{\mu \beta }F_{\alpha \nu }\right] \gamma
^{0}\sigma ^{\mu \beta }\sigma ^{\alpha \nu }$. The EDM and MDM
nonrelativistic contributions are
\begin{equation}
H_{\text{NR(4)}}=\lambda _{4}T_{0i}E^{j}\epsilon _{ijk}\sigma ^{k}+\lambda
_{4}T_{ad}B^{a}\sigma ^{d}.
\end{equation}%
In this case, there are unconventional MDM and EDM contributions. The
bounds achieved for this model are also included in Table I. It is possible
to show that many other couplings involving four gamma matrices are either
physically redundant or do not generate spin interactions.\cite{Jonas2}

In this work, we analyzed a new class of dimension-five, CPT-even LV
nonminimal couplings between fermions and photons, composed of a general
tensor, $T_{\mu \nu },$ in the context of the Dirac equation, addressing its
axial and non-axial versions. EDM and MDM experimental data were used to
impose upper bounds on the LV parameters of $1$
part in $10^{11}$ and $10^{16}$ GeV$^{-1}$. The ``rotated" MDM and EDM
unconventional interpretation could allow constraining off-diagonal
components of the LV tensor if experiments able to measure EDM and MDM
components orthogonal to the applied fields are conceived.

\section*{Acknowledgments}
The authors are grateful to the Brazilian research agencies
CNPq, CAPES, and FAPEMA for invaluable financial support.

\end{document}